\documentclass[pra,aps,twocolumn,showpacs]{revtex4}

\usepackage{bm}
\usepackage{amsmath}
\usepackage{graphicx}
\usepackage{dcolumn}
\usepackage{bm}
\begin{document}

\author{Stefan Scheel}
\email{s.scheel@imperial.ac.uk}
\author{Rachele Fermani}
\author{E.A.~Hinds}
\title{On the feasibility of studying vortex noise in 2D
superconductors with cold atoms}
\affiliation{Quantum Optics and Laser Science, Blackett Laboratory,
Imperial College London, Prince Consort Road, London SW7 2BW}
\begin{abstract}
We investigate the feasibility of using ultracold neutral atoms
trapped near a thin superconductor to study vortex noise close to
the Kosterlitz-Thouless-Berezinskii transition temperature. Alkali
atoms such as rubidium probe the magnetic field produced by the
vortices. We show that the relaxation time $T_1$ of the Zeeman
sublevel populations can be conveniently adjusted to provide long
observation times. We also show that the transverse relaxation times
$T_2$ for Zeeman coherences are ideal for studying the vortex noise.
We briefly consider the motion of atom clouds held close to the
surface as a method for monitoring the vortex motion.
\end{abstract}
\date{\today}
\maketitle

Ultracold neutral atoms trapped and manipulated on atom chips
\cite{specialissue} can be used as sensitive probes for a wide range
of phenomena. Examples include accurate measurements of gravity
\cite{gravity}, imaging the magnetic or electric landscape near
wires and magnetic films \cite{Schmiedmayer05,
Schmiedmayer06,Hall06}, measuring the field due to Johnson noise
near metallic and dielectric surfaces
\cite{Jones03,Harber03,Lin04,Sinclair05} and probing the
Casimir-Polder force as a function of atom-surface distance
\cite{Harber05}. Until recently, the atom chips used in these kinds
of measurements have been at room temperature and the corresponding
thermal fluctuations of the magnetic field have caused mixing of the
Zeeman sublevels. Depending on the material of the surface and its
distance from the atoms, this spin relaxation time is typically of
order 1-100 seconds.

The use of superconducting films at cryogenic temperatures has been
proposed as a way to reduce thermal noise \cite{Scheel05} and cold
atoms have now been trapped near a superconducting surface
\cite{Haroche06}. In theoretical studies of atom-superconductor
interactions, the arguments employed so far have been based on a
simple two-fluid model \cite{Skagerstam06} or on data extracted from
surface impedance measurements on bulk superconductors
\cite{Scheel05}. These approaches apply to three-dimensional
superconducting materials, but the thin films normally used on atom
chips are typically closer to two-dimensional objects for which 3D
theory is not strictly valid. Near thin film superconductors, the
magnetic noise is generated primarily  by vortex motion, which is
absent in 3D superconductors. This raises the possibility that atoms
trapped near the surface of a superconducting atom chip might be
able to probe the physics of vortices. In this article, we start to
explore the feasibility of using cold atoms to investigate vortex
noise in a 2D superconductor.

Rather than going into the theory of 2D superconductors and their
vortex dynamics, it is enough for our present purpose to be guided
by existing experimental data. In the experiment reported in
\cite{Shaw96}, a SQUID loop roughly $1$mm in diameter measured the
flux noise spectrum produced by vortices in a 2D superconducting
Josephson junction array some $100\mu$m away. The detected flux
provided statistical information on the dynamics of the vortices in
the Josephson array. Here we explore how the observed vortex
behaviour would influence the magnetic sublevel populations and spin
coherence of ultracold atoms trapped above such a surface.

Our first goal is to relate the flux noise power measured in
\cite{Shaw96} to the  expected spin flip rate for a magnetic atom
trapped near the surface. To this end, we write the spectral density
of the flux noise (per Hz of bandwidth) at angular frequency
$\omega$ as
\begin{equation}
\label{eq:fluxnoise} S_\phi(f) =2 \pi \int\limits_A d^2x\, d^2y \,
\langle \hat{B}_z(\mathbf{x},z;\omega)
\hat{B}^\dagger_z(\mathbf{y},z;\omega) \rangle
\end{equation}
where $A$ is the area of the pick-up loop placed parallel to the
surface at height $z$. The points $(\mathbf{x},z)$ and
$(\mathbf{y},z)$ are any two points in the plane of the loop and
$\hat{B}_z$ is the operator for the magnetic field component normal
to the loop. Both coordinates are  integrated over the the surface
of the loop to obtain the flux noise. In thermal equilibrium, the
integrand can be written in the standard way in terms of the dyadic
Green function and the mean thermal photon number
$\bar{n}_{\mathrm{th}}$:
\begin{eqnarray}
\label{eq:expectation} \lefteqn{ \langle
\hat{B}_z(\mathbf{x},z;\omega)
\hat{B}^\dagger_z(\mathbf{y},z;\omega^{\prime}) \rangle =\delta(\omega-\omega^{\prime}) \frac{\hbar\mu_0}{\pi} } \nonumber \\
&& \times \mathrm{Im}\left[ \overrightarrow{\bm{\nabla}} \times
\bm{G}(\mathbf{x},z;\mathbf{y},z;\omega) \times
\overleftarrow{\bm{\nabla}} \right]_{zz} (\bar{n}_{\mathrm{th}}+1)
\,.
\end{eqnarray}
If atoms are trapped near the superconductor with their spins
parallel to the surface, this same noise in the magnetic field
component $B_z$ can drive spin flip transitions. For an atom at
position $\mathbf{r}_A$, the rate is given by
\begin{equation}
\label{eq:fliprate} \Gamma_z = \frac{2\mu_0\mu_{12}^2}{\hbar}
\mathrm{Im}\left[ \overrightarrow{\bm{\nabla}} \times
\bm{G}(\mathbf{r}_A,\mathbf{r}_A;\omega) \times
\overleftarrow{\bm{\nabla}} \right]_{zz} (\bar{n}_{\mathrm{th}}+1)
\end{equation}
where $\mu_{12}$ is the magnetic dipole transition matrix element
between the initial and final Zeeman sublevels. Here $\omega$ is the
resonant transition frequency, corresponding to the atomic level
splitting. The total spin-flip rate $\Gamma$ is related to
$\Gamma_z$ by $\Gamma=3/2\Gamma_z$.

Neglecting the vacuum contribution to the spin flip rate and using
the Weyl expansion for the scattering part of the Green function
\cite{Fermani06,Dung98}, we find that
\begin{eqnarray}
\label{eq:weyl} \lefteqn{ \left[ \overrightarrow{\bm{\nabla}} \times
\bm{G}(\mathbf{x},z;\mathbf{y},z;\omega) \times
\overleftarrow{\bm{\nabla}} \right]_{zz} = } \nonumber \\ && \int
\frac{d^2k_\|}{(2\pi)^2}
\,e^{i\mathbf{k}_\|\cdot(\mathbf{x}-\mathbf{y})} \,\,r_{\mathrm{s}}
\,\frac{ik_\|^2}{k_z} e^{2ik_zz}
\end{eqnarray}
where the perpendicular and parallel wavevector components are
related by $k_z=\sqrt{\omega^2/c^2-k_\|^2}$ and $r_{\mathrm{s}}$ is
the Fresnel reflection coefficient for $s$-polarized (TE) waves,
whose electric vector is perpendicular to the plane of incidence.

In order to integrate over the circular pick-up loop of radius $R$,
we use the identity
\begin{equation}
\int\limits_0^{2\pi} d\varphi\,
e^{i\mathbf{k}_\|\cdot(\mathbf{x}-\mathbf{y})}
= 2\pi J_0(k_\|l) \,,
\end{equation}
where $\mathbf{k}_\|\cdot(\mathbf{x}-\mathbf{y})=k_\|l\cos\varphi$
with $l=|\mathbf{x}-\mathbf{y}|$ and $J_0(k_\|l)$ is the
zeroth-order Bessel function. For the radial integration we have
\begin{equation}
\int\limits_0^R dl\,lJ_0(k_\|l) = \frac{R}{k_\|} J_1(k_\|R) \,,
\end{equation}
giving the result
\begin{equation}
\label{eq:pickupint} \int\limits_A d^2x\, d^2y
\,e^{i\mathbf{k}_\|\cdot(\mathbf{x}-\mathbf{y})} = A^2
\frac{4}{(k_\|R)^2} J_1^2(k_\|R)
\end{equation}
for double integration over the pickup loop. Here we explicitly pull
out the factor $A^2=(\pi R^2)^2$, which is the squared area of the
loop.

The Weyl expansion (\ref{eq:weyl}) of the Green function also
requires us to integrate over transverse wave vectors. Since this
cannot be done in closed form, we express the right hand side of
Eq.\,(\ref{eq:pickupint}) as a power series in $k_\|$
\cite{Gradsteyn},
\begin{equation}
\label{eq:bessel}
\frac{4J_1^2(k_\|R)}{(k_\|R)^2}  =
\sum\limits_{s=0}^\infty
\frac{4(-1)^sR^{2s}\Gamma(s+\frac{3}{2})}
{\sqrt{\pi}\Gamma(s+1)\Gamma(s+2)\Gamma(s+3)} k_\|^{2s} \,.
\end{equation}
In cases of practical interest, the distance $z$ between the trapped
atoms and the surface (typically 1-100$\mu$m) is very small in
comparison with the free-space wavelength of the spin-flip
transition (typically 3cm-300m). As a result, the integral over
$k_\|$ is entirely dominated by the region in which
$k_\|^2\gg\omega^2/c^2$, where $k_z\approx ik_\|$. When
Eq.~(\ref{eq:weyl}) is integrated to obtain the flux, in accordance
with Eq.~(\ref{eq:fluxnoise}), the powers $k_\|^{2s}$ arising from
the expansion (\ref{eq:bessel}) of the Bessel function can be
obtained by differentiating with respect to the atom-surface
distance $z$, that is, $k_\| \equiv -\frac{1}{2}
\frac{\partial}{\partial z}$ \cite{Fermani06}. Thus, inserting
Eq.~(\ref{eq:pickupint}), together with the power series expansion
(\ref{eq:bessel}), into Eq.~(\ref{eq:fluxnoise}), we obtain
\begin{eqnarray}
\label{eq:gammal} S_\phi(f) &=& \frac{\hbar^2(\pi
R^2)^2}{\mu_{12}^2} \sum\limits_{s=0}^\infty
\frac{4(-1)^s\Gamma(s+\frac{3}{2})}
{\sqrt{\pi}\Gamma(s+1)\Gamma(s+2)\Gamma(s+3)} \nonumber \\ && \times
\left( \frac{R}{2} \right)^{2s} \frac{\partial^{2s}}{\partial
z^{2s}} \Gamma_z \,.
\end{eqnarray}

The final approximation consists of assuming that the spin flip rate
$\Gamma_z$ follows a strict power law with respect to the
atom-surface distance: $\Gamma_z\propto z^{-n}$. This is certainly
the case in various limiting regimes
\cite{Scheel05,Henkel99,Henkel01,Henkel05} when the length scales
relevant to the problem (transition wavelength, atom-surface
distance, skin depth of the substrate material etc.) can be
well-separated. The derivatives in Eq.~(\ref{eq:gammal}) then become
\begin{equation}
\frac{\partial^{2s}}{\partial z^{2s}} \Gamma_z
= \frac{(n+2s-1)!}{(n-1)!z^{2s}} \Gamma_z \,.
\end{equation}
In this way, we perform the summation over $s$ in
Eq.~(\ref{eq:gammal}) to obtain
\begin{eqnarray}
\label{eq:gammal2} \lefteqn{ S_\phi(f) =
\frac{\hbar^2A^2}{\mu_{12}^2} \Gamma_z } \nonumber \\ && \times
{}_3F_2\left[\left\{\frac{3}{2},\frac{n+1}{2},\frac{n}{2}\right\},
\{2,3\},-\frac{R^2}{z^2}\right]
\end{eqnarray}
where ${}_3F_2$ denotes a hypergeometric function.
Equation~(\ref{eq:gammal2}) is the result we were seeking,
connecting the measured flux noise spectrum $S_\phi(f)$ to the
anticipated atomic spin flip rate $\frac{3}{2}\Gamma_z$. The
dependence of this connection on distance is controlled by the
argument $(R/z)^2$ of the hypergeometric function and by the power
law associated with the spin flip rate.

In the limit of small $R/z$, when the size of the pick-up loop is
small compared to its distance from the surface, the hypergeometric
function in Eq.~(\ref{eq:gammal2}) can be approximated
\cite{Fermani06} by
\begin{eqnarray}
\label{eq:rllz} \lefteqn{
{}_3F_2\left[\left\{\frac{3}{2},\frac{n+1}{2},\frac{n}{2}\right\},
\{2,3\},-\frac{R^2}{z^2}\right] } \nonumber \\ && \stackrel{R\ll
z}{\approx} 1-\frac{n(n+1)}{16} \frac{R^2}{z^2} +{\cal
O}\left(\frac{R^4}{z^4}\right)\,.
\end{eqnarray}
Hence the flux noise spectrum is essentially proportional to the spin 
flip rate times the squared area of the pick-up loop. This
reflects the fact that the loop is small compared with the
transverse correlation length (of order $z$) and therefore the flux
directly samples the local magnetic field.

The flux measurements reported in \cite{Shaw96} were made in the
opposite limit, $R\gg z$, with a large pick-up loop located very
close to the superconducting surface.  Assuming a power law
$\Gamma_z\propto 1/z^4$ (corresponding to the limit $\delta\ll z$
with $\delta$ being the penetration depth of the substrate material
\cite{Henkel99,Henkel01,Henkel05,Scheel05}), we obtain the limiting
behaviour
\begin{eqnarray}
\label{eq:rggz} \lefteqn{
{}_3F_2\left[\left\{\frac{3}{2},\frac{5}{2},2\right\},
\{2,3\},-\frac{R^2}{z^2}\right] \equiv
{}_2F_1\left[\frac{3}{2},\frac{5}{2},3,-\frac{R^2}{z^2}\right] }
\nonumber \\ && \stackrel{R\gg z}{\approx} \frac{16}{3\pi}
\frac{z^3}{R^3} +\frac{4}{3\pi} \frac{z^5}{R^5} \left[
5+6\log\frac{z}{4R}\right] +{\cal O}\left(\frac{z^7}{R^7}\right)\,.
\end{eqnarray}
Comparing Eqs.~(\ref{eq:rggz}) and (\ref{eq:rllz}) we see that the
spectral density of the flux noise measured in a large loop is also
proportional to $A^2\Gamma_z$ but is suppressed by an additional
factor of order $(z/R)^3$.

\begin{figure}
  \includegraphics[width=8cm]{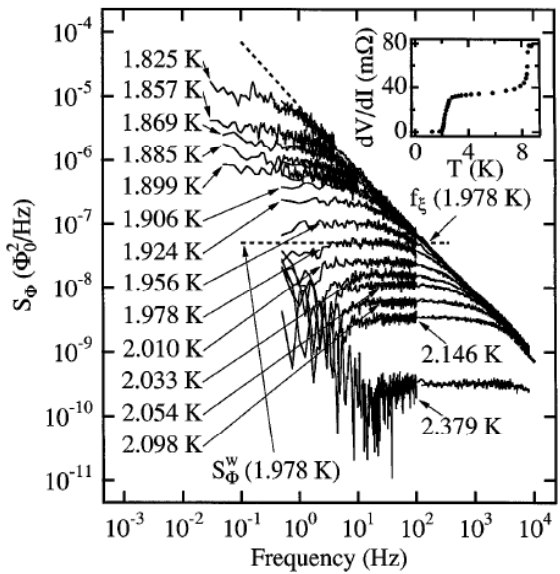}\\
  \caption{Reproduced from Ref.~\cite{Shaw96}. Spectral density of magnetic flux noise, $S_\Phi (f)$, versus
frequency $f=\omega /2 \pi$ for 15 temperatures above the
Kosterlitz-Thouless-Berezinskii transition. Scatter at higher
temperatures is due to subtraction of SQUID noise. Dashed lines have
slope $-1$ and $0$. Inset shows $dV/dI$ versus T.}\label{fig1}
\end{figure}

For the purpose of quantitative comparison, we take the magnetic
dipole matrix element to be $\mu_{12}=\mu_B/2$, corresponding to a
transition between Zeeman sublevels $|i\rangle=|F=2,m_F=2\rangle$
and $|f\rangle=|F=2,m_F=1\rangle$ of a ground-state rubidium atom.
Equation~(\ref{eq:gammal2}) then gives
\begin{equation}
S_\phi(f) = \frac{16\,m_e^2}{e^2}\;\Gamma_z A^2 f(z,R) \,,
\end{equation}
where $e$ and $m_e$ are the charge and mass of the electron and
$f(z,R)$ denotes the function in Eq.~(\ref{eq:rllz}) or
Eq.~(\ref{eq:rggz}). In order to make contact with the flux
measurements reported in Ref.~\cite{Shaw96} and reproduced in
Fig.~\ref{fig1}, we write $S_\phi(f)=x\cdot\Phi_0^2$, where
$\Phi_0=h/(2e)$ is the flux quantum. This gives
\begin{equation}
\label{eq:ratevsnoise} \Gamma_z = x\cdot
\frac{\pi^2\hbar^2}{16\,m_e^2A^2f(z,R)} \stackrel{z\ll R}{\simeq}
x\cdot \frac{3\pi^3\hbar^2}{256\,m_e^2A^2}
\left(\frac{R}{z}\right)^3 \,.
\end{equation}
The SQUID loop that measured the flux in \cite{Shaw96} had an
effective area of $A=180\times 900\mu\mathrm{m}^2\approx 2\cdot
10^{-7}\mathrm{m}^2$ and had $R/z\geq2.3$, giving a spin flip rate
of $\Gamma_z\geq x\cdot 2\times10^6~\mbox{s}^{-1}$. At spin-flip
frequencies above $10\mbox{kHz}$ (corresponding to a quantisation
magnetic field stronger than $1\,\mu\mbox{T}$), the value of $x$
given in Ref.~\cite{Shaw96} is below $10^{-9}$, corresponding to a
trap lifetime in excess of $500\,\mbox{s}$. This rather slow $T_1$
relaxation rate is very promising from the point of view of keeping
atoms trapped near a superconducting surface. At the same time it
may well be fast enough to be measured in the very benign
environment of a cryostat.

As well as inducing atomic spin flips, the magnetic field
fluctuations can generate noise in the relative phase between Zeeman
sublevels. In the presence of a static field $B_0$ normal to the
surface, the variance of the phase between levels 1 and 2 after time
$T$ is given by
\begin{equation}
\label{eq:phasevariance} \sigma \left[\phi(T)\right]^2 =
\frac{(\mu_{22}-\mu_{11})^2}{\hbar^2} \int\limits_0^T
dt\int\limits_0^T dt^\prime
\langle\hat{B_z}(t)\hat{B_z}(t^\prime)\rangle
 \,,
\end{equation}
where $\hat{B_z}(t)$ is the noise field and does not include the
constant field $B_0$. This phase noise can be related to the spin
flip rate. We connect $\hat{B_z}(t)$ to $\hat{B_z}(\omega)$ through
\begin{equation}
\label{eq:FT} \hat{B}_z(t)= \int\limits_0^\infty
d\omega\left[\hat{B}_z(\omega)e^{-i \omega t}+\mbox{h.c.}\right]
 \,,
\end{equation}
and we use Eqs.~(\ref{eq:expectation}) and (\ref{eq:fliprate}) to
obtain
\begin{equation}
\label{eq:FDT}\mu_{12}^2
\langle\hat{B}_z(\omega)\hat{B}^\dagger_z(\omega^\prime)\rangle
=\frac{\hbar^2}{2\pi}\Gamma_z(\omega)\delta(\omega-\omega^\prime)\,.
\end{equation}
Substitution of Eqs.~(\ref{eq:FT}) and (\ref{eq:FDT}) into
Eq.~(\ref{eq:phasevariance}) then yields the result
\begin{equation}
\label{eq:phaseANDgamma} \sigma[\phi(T)]^2 =
\frac{(\mu_{22}-\mu_{11})^2}{\mu_{12}^2}
\,\,\frac{2}{\pi}\int\limits_0^\infty
d\omega\,\Gamma_z(\omega)\frac{1-\cos(\omega T)}{\omega^2}\,.
\end{equation}

Now, $\Gamma_z(\omega)$ is proportional to the measured flux noise
$S_\phi(f)$ [Eq.~(\ref{eq:ratevsnoise})], which we know is constant
up to a characteristic frequency $f_\xi$, as shown in
Fig.~\ref{fig1}. Let us call this low frequency rate $\Gamma(0)$.
Above that, $\Gamma_z(\omega)$ drops off as roughly $1/\omega$. The 
integral in Eq.~(\ref{eq:phaseANDgamma}) is completely dominated by the 
low frequency range between $0\leq\omega\leq2\pi/T$, so we can make the
approximation $\Gamma_z(\omega)=\Gamma(0)$ provided the observation
time satisfies $T>1/f_\xi$. Since $f_\xi$ exceeds 100~Hz, this is
the case for $T>10\,\mbox{ms}$. Indeed, for large enough observation
times the integral kernel approximates the $\delta$ function,
\begin{equation}
\frac{2}{\pi} \frac{(1-\cos\omega T)}{\omega^2} \mapsto T \delta(\omega)
\,.
\end{equation}
Equation~(\ref{eq:phaseANDgamma}) then reads simply
\begin{equation} \label{eq:finalphase} \sigma[\phi(T)]^2 =
\frac{(\mu_{22}-\mu_{11})^2}{\mu_{12}^2} \,\,\Gamma_z(0)T.
\end{equation}

Supposing once again that states 1 and 2 are the ground states
$|F=2,m_F=2\rangle$ and $|F=2,m_F=1\rangle$ of a rubidium atom, the
ratio of magnetic matrix elements squared has the value of unity and
we obtain the particularly simple result $\sigma[\phi(T)]^2 =
\Gamma_z(0)T\simeq 2\times10^6x(0)T$. Since the value of $x(0)$
reported in \cite{Shaw96} is in the range  $10^{-9}-10^{-5}$, the
corresponding dephasing lifetime $T_2$ is in the range
$50~\mbox{ms}-500~\mbox{s}$. This provides a very convenient time
scale for the study of vortex noise using Ramsey interferometry, in
which atoms prepared in a coherent superposition of Zeeman sublevels
are later interrogated to measure the time-evolution of the
coherence. The vortex field noise would be manifest as a loss of
Ramsey fringe visibility with time, which could be studied as a
function of the atom-surface distance. It should also be possible to
explore the transverse coherence of the noise by varying the
transverse extent of the cloud and measuring the loss of Ramsey
fringe visibility as the cloud length increases.

Measurements using cold atoms may also be able to image the
vortices. The typical vortex separation of $\xi\sim 2\mu$m
\cite{Shaw96}, could be resolved by bringing the atom cloud to a
similar distance from the surface, where the field of each vortex is
of order $\Phi_0/(\pi \xi^2)\sim 1\,$G. At this close approach, the
spin-flip lifetime is strongly reduced, but lifetimes approaching
100\,ms may nevertheless be achieved. One imaging method would be to
study the density distribution of the atoms, which is altered by the
presence of the vortices through the effect of the vortex fields on
the trapping potential. This approach is used to image classical
current distributions in wires on an atom chip
\cite{Schmiedmayer05}. The motion of the vortices could be tracked
through the motion of the density patterns in the atom cloud.

To conclude, we have shown that it is feasible to detect vortex
dynamics in two-dimensional superconducting films by means of
trapped cold  neutral atoms. In particular we have considered the
rate of atomic spin flips due to the magnetic field noise from the
vortex motion. At $100\,\mu$m from the surface we find that this
lifetime can be in excess of 500\,s, giving ample time to study the
vortices. We have also considered the dephasing time for
superpositions of Zeeman sublevels and find that this is short
enough to be a sensitive measure of the vortex field noise. Finally,
we have noted that spatial imaging of the vortices should be
possible using cold atoms trapped close enough to the surface. These
estimates show that cold atom clouds offer a sensitive new probe for
the study of vortex dynamics in superconducting thin films.

This work was supported by the European Commission through the Atom
Chips, Conquest and SCALA networks and by the UK through EPSRC,
QIP IRC and Royal Society funding. We acknowledge discussions with
A.~Armour, K.~Benedict and J.~Fort\'{a}gh.

\end{document}